\documentclass[journal,comsoc]{IEEEtran}
\usepackage{array}
\usepackage{url}
\usepackage{color}
\RequirePackage{filecontents}
\usepackage{cite}
\usepackage{graphicx}
\usepackage{psfrag}
\usepackage{paralist}
\usepackage{subcaption}
 \usepackage{multirow}
 \usepackage[table,xcdraw]{xcolor}
\usepackage[acronym]{glossaries}
\usepackage{amsmath,stackrel}
\usepackage{amsfonts}
\usepackage{mathalfa}
\usepackage{amssymb,mathtools}
\usepackage{amscd}
\usepackage{tikz}
\usepackage{circuitikz}
\usetikzlibrary{shapes.geometric,arrows,positioning,patterns,matrix}
\usepackage{pgfplots}
\usepackage{pgfplotstable}
\usepgfplotslibrary{patchplots}
\pgfplotsset{compat=1.13}
\usepackage{tabstackengine} 

\usepackage{verbatim}
\usepackage{enumerate}                     
\usepackage{amsthm}
\usepackage{eucal}
\usepackage{algorithmic}
\usepackage{algorithm}
\usepackage[nolist]{acronym}  
\usepackage[normalem]{ulem}

\newcommand{\EX}[1]  {{\mathbb{E}}\left\{{#1}\right\}}

\theoremstyle{plain}

\theoremstyle{plain}

\theoremstyle{plain}

\theoremstyle{plain}

\theoremstyle{remark}
\newtheorem{remark}{Remark}
\theoremstyle{remark}

\theoremstyle{definition}

\theoremstyle{plain}
\newtheorem{property}{Property}

\begin{document}
\begin{acronym}
	\acro{lora}[LoRa]{low power long range}
	\acro{lorawan}[LoRaWAN]{Low power, Long Range Wide Area Network}
	\acro{lpwan}[LPWAN]{Low Power Wide Area Network}
	\acro{ed}[ED]{end device}
	\acro{ns}[NS]{network server}
	\acro{gw}[GW]{gateway}
	\acro{fsk}[FSK]{frequency shift key}
	\acro{sf}[SF]{spreading factor}
	\acro{ISM}{Industrial, Scientific and Medical}
	\acro{ETSI}{European telecommunications standards institute}
	\acro{CPS}{cyber-physical systems}
	\acro{CPScapital}[CPS]{Cyber-physical systems}
	\acro{AC}{address coding}
	\acro{ACF}{autocorrelation function}
	\acro{ACR}{autocorrelation receiver}
	\acro{ADC}{analog-to-digital converter}
	\acrodef{aic}[AIC]{Analog-to-Information Converter}     
	\acro{AIC}[AIC]{Akaike information criterion}
	\acro{aric}[ARIC]{asymmetric restricted isometry constant}
	\acro{arip}[ARIP]{asymmetric restricted isometry property}
	
	\acro{ARQ}{automatic repeat request}
	\acro{AUB}{asymptotic union bound}
	\acrodef{awgn}[AWGN]{Additive White Gaussian Noise}     
	\acro{AWGN}{additive white Gaussian noise}
	\acro{waric}[AWRICs]{asymmetric weak restricted isometry constants}
	\acro{warip}[AWRIP]{asymmetric weak restricted isometry property}
	\acro{BCH}{Bose, Chaudhuri, and Hocquenghem}        
	\acro{BCHC}[BCHSC]{BCH based source coding}
	\acro{BEP}{bit error probability}
	\acro{BFC}{block fading channel}
	\acro{BG}[BG]{Bernoulli-Gaussian}
	\acro{BGG}{Bernoulli-Generalized Gaussian}
	\acro{BPAM}{binary pulse amplitude modulation}
	\acro{BPDN}{Basis Pursuit Denoising}
	\acro{BPPM}{binary pulse position modulation}
	\acro{BPSK}{binary phase shift keying}
	\acro{BPZF}{bandpass zonal filter}
	\acro{BSC}{binary symmetric channels}              
	\acro{BU}[BU]{Bernoulli-uniform}
	
	\acro{CDF}{cumulative distribution function}
	\acro{CCDF}{complementary cumulative distribution function}
	\acro{CD}{cooperative diversity}
	
	\acro{CDMA}{code division multiple access}
	\acro{ch.f.}{characteristic function}
	\acro{CIR}{channel impulse response}
	\acro{cosamp}[CoSaMP]{compressive sampling matching pursuit}
	\acro{CR}{cognitive radio}
	\acrodef{cscapital}[CS]{Compressed sensing}
	\acro{CS}[CS]{compressed sensing}
	\acro{CSI}{channel state information}

	\acro{DAA}{detect and avoid}
	\acro{DAB}{digital audio broadcasting}
	\acro{DCT}{discrete cosine transform}
	\acro{dft}[DFT]{discrete Fourier transform}
	\acro{DR}{distortion-rate}
	\acro{DS}{direct sequence}
	\acro{DS-SS}{direct-sequence spread-spectrum}
	\acro{DTR}{differential transmitted-reference}
	\acro{DVB-H}{digital video broadcasting\,--\,handheld}
	\acro{DVB-T}{digital video broadcasting\,--\,terrestrial}

	\acro{ECC}{European Community Commission}
	\acro{EED}[EED]{exact eigenvalues distribution}
	\acro{ELP}{equivalent low-pass}

	\acro{FC}[FC]{fusion center}
	\acro{FCC}{Federal Communications Commission}
	\acro{FEC}{forward error correction}
	\acro{FFT}{fast Fourier transform}
	\acro{FH}{frequency-hopping}
	\acro{FH-SS}{frequency-hopping spread-spectrum}
	\acrodef{FS}{Frame synchronization}
	\acro{FSK}{Frequency Shift Key}
	\acro{FSsmall}[FS]{frame synchronization}

	\acro{GA}{Gaussian approximation}
	\acro{GF}{Galois field }
	\acro{GG}{Generalized-Gaussian}
	\acro{GIC}[GIC]{generalized information criterion}
	\acro{GLRT}{generalized likelihood ratio test}
	\acro{GPS}{Global Positioning System}

	\acro{i.i.d.}{independent, identically distributed}
	\acro{IoT}{Internet of Things}

	\acro{LF}{likelihood function}
	\acro{LLF}{log-likelihood function}
	\acro{LLR}{log-likelihood ratio}
	\acro{LLRT}{log-likelihood ratio test}
	\acro{LOS}{line-of-sight}
	\acro{LRT}{likelihood ratio test}

	\acro{MB}{multiband}
	\acro{MC}{multicarrier}
	\acro{MDS}{mixed distributed source}
	\acro{MF}{matched filter}
	\acro{m.g.f.}{moment generating function}
	\acro{MI}{mutual information}
	\acro{MIMO}{multiple-input multiple-output}
	\acro{MISO}{multiple-input single-output}
	\acrodef{maxs}[MJSO]{maximum joint support cardinality}                                           
	\acro{ML}[ML]{maximum likelihood}
	\acro{MMSE}{minimum mean-square error}
	\acro{MMV}{multiple measurement vectors}
	\acrodef{MOS}{model order selection}
	\acro{M-PSK}[${M}$-PSK]{$M$-ary phase shift keying}                        
	\acro{M-QAM}[$M$-QAM]{$M$-ary quadrature amplitude modulation}             
	\acro{MRC}{maximal ratio combiner}                  
	\acro{maxs}[MSO]{maximum sparsity order}                                    
	\acro{M2M}{machine to machine}                                                
	\acro{MUI}{multi-user interference}

	\acro{NB}{narrowband}
	\acro{NBI}{narrowband interference}
	\acro{NLA}{nonlinear sparse approximation}
	\acro{NLOS}{non-line-of-sight}
	\acro{NTIA}{National Telecommunications and Information Administration}

	\acro{PAM}{pulse amplitude modulation}
	\acro{PAR}{peak-to-average ratio}
	\acrodef{pdf}[PDF]{probability density function}                      
	\acro{PDF}{probability density function}
	\acrodef{p.d.f.}[PDF]{probability distribution function}
	\acro{PDP}{power dispersion profile}
	\acro{PMF}{probability mass function}                              
	\acrodef{p.m.f.}[PMF]{probability mass function}
	\acro{PN}{pseudo-noise}
	\acro{PPM}{pulse position modulation}
	\acro{PRake}{Partial Rake}
	\acro{PSD}{power spectral density}
	\acro{PSEP}{pairwise synchronization error probability}
	\acro{PSK}{phase shift keying}
	\acro{8-PSK}[$8$-PSK]{$8$-phase shift keying}
	\acro{QAM}{quadrature amplitude modulation}
	\acro{QPSK}{quadrature phase shift keying}		
	\acro{RD}[RD]{raw data}
	\acro{RDL}{"random data limit"}
	\acro{ric}[RIC]{restricted isometry constant}
	\acro{rict}[RICt]{restricted isometry constant threshold}
	\acro{rip}[RIP]{restricted isometry property}
	\acro{ROC}{receiver operating characteristic}
	\acro{rq}[RQ]{Raleigh quotient}
	\acro{RS}[RS]{Reed-Solomon}
	\acro{RSC}[RSSC]{RS based source coding}
	\acro{r.v.}{random variable}                                
	\acro{R.V.}{random vector}	
	\acro{WSN}{wireless sensor network}                        
	
\end{acronym}
\def\tsa{T}
\def\T{T_{s}}
\def\L{L}
\def\k{k}
\def\al{a_{\ell}}
\def\anr{{\mathrm{a}}_{n}}
\def\an{{{A}}_{n}}
\def\abn{{{\bf A}}_{n}}
\def\a{a}
\def\ta{\tau_{a}}
\def\ak{a_{k}}
\def\l{\ell}
\def\SF{\text{SF}}
\title{On the LoRa Modulation for IoT: Waveform Properties and Spectral Analysis}
\author{Marco~Chiani,~\IEEEmembership{Fellow,~IEEE}, and Ahmed~Elzanaty,~\IEEEmembership{Member,~IEEE} 
\thanks{M.~Chiani is with the Department of Electrical, Electronic and Information Engineering ``G. Marconi'' (DEI), University of Bologna,
Via dell'Universit\`a 50, Cesena, Italy (e-mail: marco.chiani@unibo.it).}
\thanks{A. Elzanaty was with the Department of Electrical, Electronic and Information Engineering, University of Bologna, 40136 Bologna, Italy.
He is now with the Computer, Electrical and Mathematical Science and Engineering Division, King Abdullah University of Science and Technology, Thuwal 23955, Saudi Arabia (e-mail: ahmed.elzanaty@kaust.edu.sa).}%
\thanks{This work was supported in part by MIUR under the program ``Dipartimenti di Eccellenza (2018-2022)," and in part by the EU project eCircular (EIT Climate-KIC).}
\thanks{Copyright (c) 2019 IEEE. Personal use of this material is permitted. However, permission to use this material for any other purposes must be obtained from the IEEE by sending a request to pubs-permissions@ieee.org.}
}
\markboth{ACCEPTED FOR PUBLICATION in IEEE Internet of Things Journal}{On the LoRa Modulation for IoT: Waveform Properties and Spectral Analysis}
\maketitle

\begin{abstract}
An important modulation technique for \ac{IoT} is the one proposed by the \acs{lora} alliance\texttrademark. 
In this paper we analyze the $M$-ary \acs{lora} modulation in the time and frequency domains. 
First, we provide the signal description in the time domain, and show that \acs{lora} is a memoryless continuous phase modulation. The cross-correlation between the transmitted waveforms is determined, proving that \acs{lora} can be considered approximately an orthogonal modulation only for large $M$. 
Then, we investigate the spectral characteristics of the signal modulated by random data, obtaining a closed-form expression of the spectrum in terms of Fresnel functions. 
Quite surprisingly, we found that \acs{lora} has both continuous and discrete spectra, with the discrete spectrum containing exactly a fraction $1/M$ of the total signal power. 
\end{abstract}
\section{Introduction}
The most typical \ac{IoT} scenario involves devices with limited energy, that need to be connected to the Internet via wireless links.  In this regard, \acp{lpwan} aim to offer low data rate communication capabilities over ranges of several kilometers \cite{CenVanZanZor:16,RazKulSoo:17,AdeVil:17,Mormam:17}. 
Among the current communication systems, that proposed by the \acsu{lora} alliance (Low power long Range) \cite{LoRaalliance} is one of the most promising, with an increasing number of \ac{IoT} applications, including smart metering, smart grid, and data collection from \aclp{WSN} for environmental monitoring \cite{Sherazi2018,ZhaLin:17,LeeKe:18,FerBruVer:18,PasBurFel:18,AlaPer:18}. Several works discuss the suitability of the LoRa communication system when the number of \ac{IoT} devices increases \cite{GeoRaz:2017,AbeHaxi:17,LimHan:18,CroGuc:18}. 

The modulation used by LoRa, related to Chirp Spread Spectrum, has been originally defined by its instantaneous frequency \cite{Sfo:13}. Few recent papers attempted to provide a description of the \ac{lora} modulation in the time domain, but, as will be detailed below, 
they are not complaint with the original LoRa  signal model. 
The LoRa performance has been analyzed by simulation or by considering it as an orthogonal modulation \cite{RenPol:16,Van:17,ElsRob:18}.  
On the other hand, the spectral characteristics of LoRa have not been addressed in the literature. 

{
In this paper we provide a complete characterization of the LoRa modulated signal. In particular, we start by developing a mathematical model for the modulated signal in the time domain. The waveforms of this $M$-ary modulation technique are not orthogonal, and the loss in performance with respect to an orthogonal modulation is quantified by studying their cross-correlation. The characterization in the frequency domain is given in terms of the power spectrum, where both the continuous and discrete parts are derived. The found analytical expressions are compared with the spectrum of LoRa obtained by  experimental data. 

The main contributions of this paper can be summarized as follows: 
\begin{itemize}
\item we provide the analytical expression of the signal for the $M$-ary LoRa chirp modulation in the time domain (both continuous-time and discrete-time);

\item we derive the cross-correlation between the LoRa waveforms, and prove that the modulation is non-orthogonal; 


\item we prove that the waveforms are asymptotically orthogonal for increasingly large $M$;

\item we derive explicit closed-form expressions of the continuous and discrete spectra of the LoRa signal in terms of the Fresnel functions;

\item we prove that the power of the discrete spectrum is exactly a fraction $1/M$ of the overall signal power;

\item we compare the analytical expression of the spectrum with experimental data from commercial LoRa devices;

\item we show how the analytical expressions of the spectrum can be used to investigate the compliance of the LoRa modulation with the spectral masks regulating the out-of-band emissions and the power spectral density. 

\end{itemize}


The provided time and spectral characterization of the LoRa signal is an analytical tool for the system design, as it allows suitable selection of the modulation parameters in order  to fulfill the given requirements. For example, our analysis clarifies how the spreading factor, maximum frequency deviation, and  transmitted power determine the occupied bandwidth, shape of the power spectrum and its compliance with spectrum regulations, system spectral efficiency,  total discrete spectrum power, maximum cross-correlation,  and SNR penalty with respect to orthogonal modulations. 


}

Throughout the manuscript, we define the indicator function $g_T(t)=1$ for $0\leq t <T$ and $g_T(t)=0$ elsewhere, and indicate as $u(t)$ the unit step function. The Dirac's delta is indicated as $\delta(x)$, and its discrete version as $\delta_m$, with $\delta_0=1$, $\delta_m=0 \, \forall m\ne 0$.  We also indicate with  $C(x)\triangleq \int_{0}^{x} \cos\left(t^{2} {\pi}/{2} \right) \, dt$ and $S(x) \triangleq \int_{0}^{x} \sin\left(t^{2} {\pi}/{2}\right) \, dt\,$ the Fresnel functions\cite{AbrSte:B70}.
%
\section{LoRa Signal Model}\label{sec:signalmodel}
The LoRa frequency shift chirp spread spectrum modulation has been originally described in terms of the instantaneous frequency reported in \cite[Figure 7]{Sfo:13}. 
It is an $M$-ary digital modulation, where the $M$ possible waveforms at the output of the modulator are chirp modulated signals over the frequency interval $(f_0-B/2, f_0+B/2)$ with $M$ different initial frequencies. {The data modulated signal is usually preceded by synchronization waveforms, not considered here.} 
For the data, the instantaneous frequency is linearly increased, and then wrapped to $f_0-B/2$ when it reaches the maximum frequency $f_0+B/2$, an operation that mathematically can be seen as a reduction modulo $B$. {Having the instantaneous frequency sweeping over $B$ does not imply that the signal  bandwidth is $B$, as will be discussed in Section~\ref{sec:psd}.}  

For LoRa the parameters are chosen such that $M=2^{\SF}$ with $\SF$ integer, and $B \T=M$, where $\T$ is the symbol interval.  
The bit-rate of the modulation is  
\begin{align*}
R_b&=\frac{1}{\T} \log_2 M = \frac{\SF}{\T}= B \frac{\SF}{2^{\SF}}  
\label{eq:Br}
\end{align*}
The ratio between the chip-rate $R_c=M/\T=B$ and the bit-rate is therefore\footnote{In spread-spectrum literature this is what is usually called spreading factor. However, in the LoRa terminology $\SF$ is called the spreading factor. }  
%
\begin{equation*}
\eta=\frac{R_c}{R_b}=\frac{B}{R_b}=\frac{2^{\SF}}{\SF} \,.
\label{eq:sprefac}
\end{equation*}
Its reciprocal $1/\eta$ can be seen as the modulation spectral efficiency in $\text{bit/s/Hz}$. Some values of the spectral efficiency are reported in Table~\ref{table:bw} for $M$ ranging from $2^{3}$ to $2^{12}$. 


\subsection{Continuous-time description}
\begin{figure}
	\centering
	\includegraphics[width=0.99\linewidth]{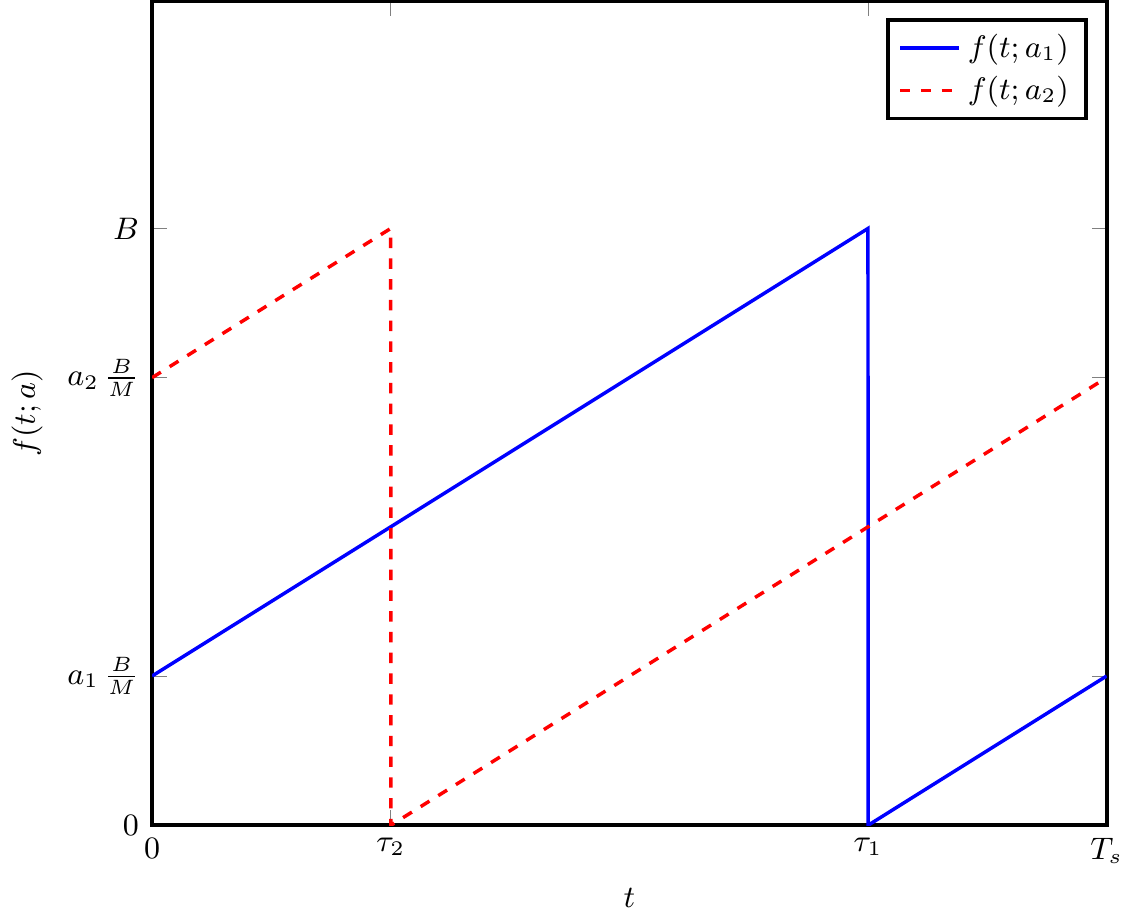}
	\caption{Example of the instantaneous frequency $f(t;\a)$ as a function of time for two different modulating symbols $a_1, a_2 \in \{0, \ldots , M-1\}$.}
	\label{fig.instfreq}
\end{figure}

\noindent To describe mathematically the signal in the time domain, let us start for clarity by assuming that the frequency interval over which to linearly sweep the frequency is $[0, B]$ as depicted in Fig.~\ref{fig.instfreq}. 
%
For the time interval $t \in [0,\T[$ and a symbol $\a \in \{0, 1, \ldots, M-1\}$ the instantaneous frequency in LoRa can thus be written as
\begin{align}\label{eq:fta}
f(t;\a)&=\a\, \frac{B}{M}+\frac{B}{\T}\,t  \pmod B \nonumber \\
&= \a\, \frac{B}{M}+\frac{B}{\T}\,t-B\, u\left(t-\ta\right) && 0 \leq t < \T   
\end{align}
where $\a\, {B}/{M}$ is the initial frequency which depends on the modulating symbol, and
\begin{align}\label{eq:ta1}
    \ta=\T\left(1-\frac{\a}{M}\right)
\end{align}
is the time instant where, after a linear increase, the instantaneous frequency reaches the maximum; for the remaining part of the symbol interval the instantaneous frequency is still linearly increasing, but reduced modulo $B$ by subtracting $B$. 

Assuming the modulation starts at $t=0$, from \eqref{eq:fta} the phase $\phi(t;\a)$ for $t\in[0,\T[$ is given by 
\begin{align}
\phi(t;\a) &\triangleq 2\pi  \int_{0}^{t} f(\tau,  \a)   \, d\tau \nonumber \\
&= 2\,\pi \left[\a\, \frac{B}{M} \,t+\frac{B}{2\,\T}\,t^2-B\,\left(t-\ta\right)\, u (t-\ta) \right] \,. 
\label{eq.phaseatt}
\end{align}
Also, with the LoRa parameters we see from \eqref{eq:ta1} that the product $B\ta=M-a$ is an integer, and can therefore be omitted in the phase. 

\begin{figure}[t!]
	\centering
	\includegraphics[width=0.75\linewidth]{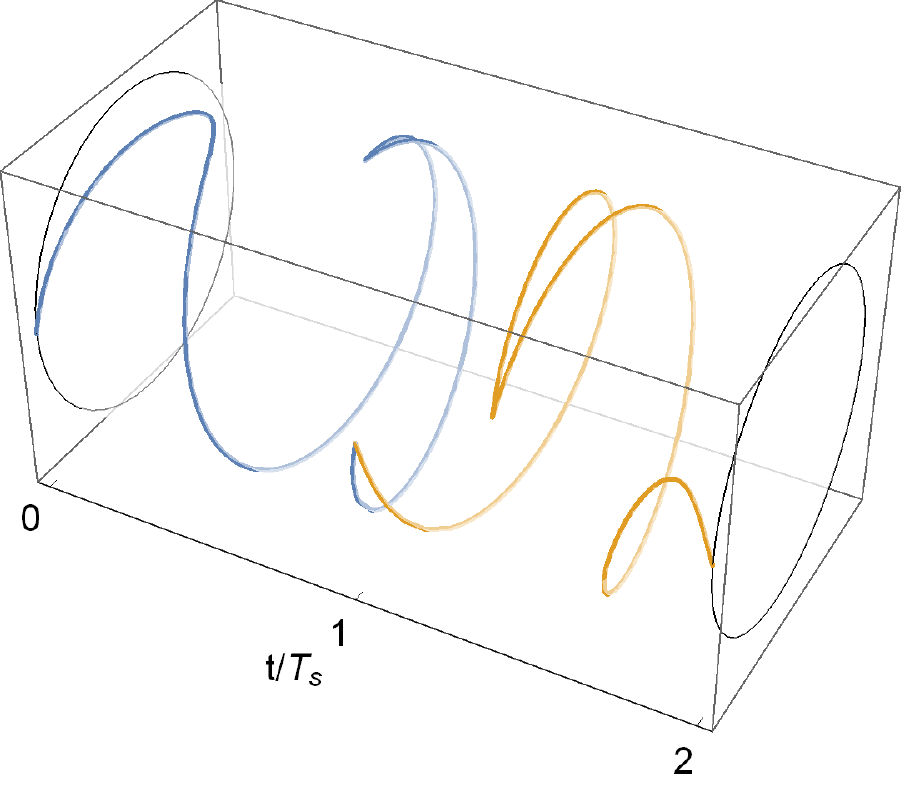}
	\caption{The phase diagram as a function of time over two consecutive LoRa modulated symbols, indicated in blue and orange.}
	\label{fig:phasediagramsf10}
\end{figure}

{Note that a factor $1/2$ for the quadratic term is missing in the phase definitions reported in \cite{RenPol:16,Van:17,ElsRob:18}, making the instantaneous frequency of the signal not complaint with that of LoRa. That difference also propagated in the discrete-time version of the signals used in \cite{Van:17,ElsRob:18}, so that even the time-discrete analysis made there is not applicable to the LoRa signal.}
\begin{property} The LoRa modulation is a memoryless continuous phase modulation with $\phi(0;\a)=\phi(\T;\a)$.
\end{property}
\begin{proof}
\noindent The initial phase is $\phi(0;\a)=0$. The phase at the end of the symbol interval is 
\begin{align}
\phi(\T;\a)&=2\,\pi \left[\a\, \frac{B}{M} \T+\frac{B}{2}\T-B \left(\T-\ta\right)\, u (\T-\ta) \right] \nonumber \\
&=2\,\pi \left(\a+\frac{M}{2}-M\, u (\T-\ta)\right) = 0 \pmod {2\pi} \nonumber
\end{align}
%
%
where the last equality is due to that $\a+{M}/{2}-M u (\T-\ta)$ is always an integer. In other words, the initial and final phases are coincident, irrespectively on the symbol $\a$. 
\end{proof}
From this property we see that the LoRa modulation can be interpreted as a continuous phase memoryless modulation, where the transmitted waveform in each symbol interval  depends only on the symbol in that interval, and not on previous or successive symbols. This can be visualized through the phase diagram which tracks the evolution of the phase over time.  In Fig.~\ref{fig:phasediagramsf10}, the phase diagram for two consecutive LoRa modulated symbols is shown as a function of time. It can be noted that each waveform starts and ends with the same phase.

The complex envelope of the modulated signal is
\begin{align}
 x(t;\a)&=\gamma \exp\left\{j\, \phi(t;\a)\right\},  && 0 \leq t < \T   
\end{align}
where $\gamma=\sqrt{2\, P_{\text s}}$ accounts for the passband signal power $P_{\text s}$. In the following we will assume $\gamma=1$ unless otherwise stated.
By introducing a frequency shift $-B /2$, the complex envelope centered at frequency zero for the interval $[0,\T[$ is 
\begin{align}
\!\!\!\!x(t;\a)&= \exp\left\{j 2 \pi B t  \left[\frac{\a}{M}\!-\!\frac{1}{2}\!+\!\frac{B t}{2 M}\! -\! u\left(t-\frac{M-a}{B}\right) \right]\right\} \,. 
\label{eq:x2}
\end{align}
Due to the memoryless nature of the modulation, the complex envelope of the LoRa signal can  be written as
\begin{align}
i(t)&=\sum_n x(t- n\T;\a_n) g_{\T}(t-n\T) 
\label{eq:itLORA}
\end{align}
where $\a_n$ is the symbol transmitted in the time interval $[n\T, (n+1)\T[$. 
We remark that, as this is a frequency modulated signal, we have $|i(t)|=1$ and the power of the signal $i(t)$ is one. 
The passband modulated signal centered at $f_0$ is then $s(t)=\Re\left\{i(t) e^{j 2 \pi f_0 t}\right\}$. 
%
%
%
%
{\renewcommand{\arraystretch}{1.5}%
\begin{table}[]
		\caption{Spectral efficiency,  maximum cross-correlation,  $99\%$-power bandwidth, total discrete spectrum power, and maximum SNR penalty.}
	\centering
	\begin{tabular}{|c|c|c|c|c|c|}
		\hline
		$M=2^\SF$    & $1/\eta$   & $\underset{\l \ne m}{\max}  \big|\Re\left\{C_{\l,m}\right\}\big|$ & $B_{99}$         & $P_\textrm{d}$   & {$\Delta_{\max}$}\\  
		      &  [bps/Hz]     &           &                     &           & {[dB]}            \\ \hline
		$2^3$       & $0.375$  &    $0.212$           &          $1.500\, B$             & $12.5 \%$           & {$1.04$}             \\ \hline
		$2^5$       & $0.156$  &     $0.091$          &          $1.185\, B$             & $3.125 \%$            & {$0.41$}           \\ \hline
		$2^7$       & $0.055$  &     $0.045$          &           $1.045 \, B$            & $0.781\%$            & {$0.20$}            \\ \hline
		$2^{10}$     & $0.0098$  &     $0.015$          &           $0.990 \, B$            & $0.098\%$           & {$0.07$}             \\ \hline
		$2^{12}$     &	$0.00293$  &	$ 0.0075$	&           $0.986\, B$ 			 	& $0.024\%$       & {$0.03$}                    \\ \hline
	\end{tabular}
	\label{table:bw}
\end{table}
\begin{property} \label{pro:xcorr}
The cross-correlation between the continuous time waveforms $x(t; \l)$ and $x(t; m)$ with $\l \ne m$ is  
\begin{align}\label{eq:xcorr}
C_{\l,m}&=\frac{1}{\T} \int_{0}^{\T} x(t; \l) x^*(t; m) dt= \nonumber \\ 
&=M \frac{e^{j 2 \pi \l (m-\l)/M}-e^{j 2 \pi m (m-\l)/M}}{j 2 \pi (M-|m-\l |) |m-\l | } 
\end{align}
and $C_{\l,\l}=1$. 
It follows that the waveforms $x(t; \l)$ and $x(t; m)$ are orthogonal (i.e, $C_{\l,m}=0$) only for $|m-\l|=2^{(p+\SF)/2}$ with $p\geq 0$ an odd (even) integer for odd (even) $\SF$. 
%
%

\noindent Moreover, since 
\begin{align}\label{eq:xcorrRe}
\Re \left\{C_{\l,m}\right\}
&=M \frac{\sin \left(\frac{2 \pi \l (m-\l)}{M}\right)-\sin \left(\frac{2 \pi m (m-\l)}{M}\right)}{2 \pi (M-|m-\l |) |m-\l | } 
\end{align}
we have that the passband waveforms $\Re \left\{x(t; \l) e^{j 2 \pi f_0 t}\right\}$ and $\Re \left\{x(t; m) e^{j 2 \pi f_0 t}\right\}$ are orthogonal (i.e, $\Re \left\{C_{\l,m}\right\}=0$)  only when $(m-\l)^2/M$ is an integer, or when $(m^2-\l^2)/M -1/2$ is an integer. \footnote{We assume $f_0 \gg B$ so that the passband waveforms are orthogonal when $\Re \left\{C_{\l,m}\right\}=0$.}

\noindent Also,  the maximum cross-correlation can be upper bounded as
\begin{equation}
\max_{\l \ne m} \left|\Re\left\{C_{\l,m}\right\} \right| \leq \max_{\l \ne m} \left|C_{\l,m}\right| \leq \frac{1}{\sqrt{2\,M}-1}.
\end{equation}
Hence, the waveforms are asymptotically orthogonal for increasing $M$:
\begin{align}\label{eq:asympcorr}
\lim_{M \to \infty} \left| \left\{C_{\l,m}\right\}\right|  = \delta_{\l-m} \,.
\end{align}
\end{property}
\noindent
\begin{proof}
The crosscorrelation between the continuous time waveforms $x(t; \l)$ and $x(t; m)$ with $\l \ne m$ and $\l >m$ can be written as 
\begin{align}
C _ { \ell , m }
&=\frac { 1 } {\T } \int _ { 0 } ^ {\T } e^{j 2 \pi \frac{B}{M}(\ell-m)\,t - B\, t\, \left[u\left(t-\tau_{\ell}\right)-u\left(t-\tau_{m}\right) \right] }\, d t \nonumber \\
&=\underbrace{\frac { 1 } {\T } \int _ { 0 } ^ {\T } e^{j 2 \pi \frac{B}{M}(\ell-m)t} \,dt}_{0} -{\frac { 1 } {\T } \int _ { \tau_{\ell} } ^ {\tau_{m}} e^{j 2 \pi \frac{B}{M}(\ell-m)t}\, dt} \nonumber \\
&\phantom{=}+ { \frac { 1 } {\T } \int _ { \tau_{\ell} } ^ {\tau_{m} } e^{j 2 \pi\left[\frac{B}{M}(\ell-m)\,t-B\,t  \right]}\,dt} \nonumber \\
&= \frac{1}{j 2 \pi (\ell-m)} \left[e^{j 2 \pi (M-\l)(\l-m)/M}-e^{j 2 \pi (M-m)(\l-m)/M}   \right]                  \nonumber\\
&\phantom{=}+\!\frac{1}{j 2 \pi (M\!+m\!-\ell)}   \left[\!e^{j 2 \pi \frac{(\ell-M)(M+m-\ell)}{M}}\!- e^{j 2 \pi \frac{(m-M)(M+m-\ell)}{M}}\!  \right] \nonumber
\end{align}
Noting the periodicity of the complex exponential function, we have
\begin{align}
C _ { \ell , m }&=\frac{1}{j 2 \pi (\ell-m)} \left[e^{j 2 \pi \l(m-\l)/M}- e^{j 2 \pi m(m-\l)/M}  \right] \nonumber\\
&\phantom{=}+\frac{1}{j 2 \pi (M+m-\ell)} \left[e^{j 2 \pi \l(m-\l)/M}-e^{j 2 \pi m(m-\l)/M}\right] \nonumber \\
&=\frac{e^{j 2 \pi \ell\frac{m-\ell}{M}}- e^{j 2 \pi m\frac{m-\ell}{M}}}{j 2 \pi} \left(\frac{1}{M+m-\l}+\frac{1}{\l-m} \right).  \label{eq:crosscorrllargerm}
\end{align}

Similarly, for  $m>\l$ we have
\begin{align}\label{eq:crosscorrmlargerl}
C _ { \ell , m }&=\frac{e^{j 2 \pi \ell\frac{m-\ell}{M}}- e^{j 2 \pi m\frac{m-\ell}{M}}}{j 2 \pi} \left(\frac{1}{M+\l-m}+\frac{1}{m-\l}       \right).
\end{align}
Putting together \eqref{eq:crosscorrllargerm} and \eqref{eq:crosscorrmlargerl}, the complex crosscorrelation, $C _ { \ell , m }$, can be derived as in \eqref{eq:xcorr}. 
The correlation in \eqref{eq:xcorr} can be zero only if the two exponentials are equal, that requires $\l (m-\l)/M= m (m-\l)/M -k$, with $k$ an integer. Thus, it must be $|m-\l|=\sqrt{k M}$. Since this must be an integer, and $M=2^\SF$, it follows that $k=2^p$ with $p\geq 0$ an odd (even) integer for odd (even) $\SF$. 

The real cross-correlation \eqref{eq:xcorrRe} follows directly, and the conditions for its zeros are straightforward observing that $\sin \alpha = \sin \beta$ for $\alpha=\beta + k 2 \pi$ or $\alpha=\pi-\beta + k 2 \pi$. 

%
 %
 
 In order to find the asymptotic behavior of the complex cross-correlation, we start by upper bounding its absolute value for $\l \ne m$. From \eqref{eq:xcorr} we have
 \begin{equation*}
 C_{\l,m}=M e^{j 2 \pi  \left(m^2-\l^2\right)/M} \, \frac{e^{-j  \pi (m-\l)^2/M}-e^{j  \pi (m-\l)^2/M}}{j 2 \pi (M-|m-\l |) |m-\l | } 
  \end{equation*}
  and therefore
  \begin{equation*}
 \left|C_{\l,m}\right|=  M  \frac{\left|\sin\left(\pi (m-\l)^2/M\right)\right|}{ \pi (M-|m-\l |) |m-\l | }\,. 
 \end{equation*}
 The first maximum for $\left|C_{\l,m}\right|$ is in the interval ${1\leq |m-\l | \leq \left\lfloor{\sqrt{M/2}}\right\rfloor}$. This is due to the following reasons:
 \begin{itemize}
 	\item $\left|C_{\l,m}\right|$ is symmetric around $M/2$;
 	\item the denominator is monotonically increasing for  ${1\leq |m-\l | \leq M/2}$;
 	\item the numerator is monotonically increasing for $1\leq |m-\l | \leq \left\lfloor{\sqrt{M/2}}\right \rfloor$, and  starts to decrease after $\left\lfloor{\sqrt{M/2}}\right\rfloor$.
 \end{itemize}
  Hence, we have
 \begin{align*}
\max_{\l \ne m} \left|C_{\l,m}\right| &=\max_{1 \leq |m-\l |\leq \left\lfloor{\sqrt{M/2}}\right\rfloor }   M  \frac{\sin\left(\pi (m-\l)^2/M\right)}{ \pi (M-|m-\l |) |m-\l | } \\
								      &\leq   \max_{1 \leq |m-\l |\leq \left\lfloor{\sqrt{M/2}}\right\rfloor } \frac{ |m-\l |}{  M-|m-\l |}
								       \\
								      &\leq \frac{1}{\sqrt{2\,M}-1}
 \end{align*}
where for the first inequality  $\sin(x) \leq x$ for $0 \leq x \leq \pi/2$ is used. For the second inequality, it is noticed that the function is  increasing in $ |m-\l |$, so its maximum value is obtained with 
$|m-\l |=\left\lfloor{\sqrt{M/2}}\right\rfloor \leq \sqrt{M/2}$. Finally, taking the limit when $M\rightarrow \infty$ gives \eqref{eq:asympcorr}.
\end{proof}

{The correlation among the waveforms of the LoRa modulation has an impact on the error performance for the optimum coherent receiver over AWGN channels \cite{Ben:99,Proakis:95}. 
In particular, for the pairwise error probability between the $\l$-th and $m$-th  waveforms there is a factor $1-\Re\left\{C_{\l,m}\right\}$ in the SNR with respect to orthogonal modulation schemes (see, e.g., equations (4.31) and (4.49) in \cite{Ben:99}). 
%
	In Table~\ref{table:bw} we report the maximum penalty on the SNR, $\Delta_{\max}$, corresponding to the maximum cross-correlation, to be paid with respect to orthogonal modulation schemes. 
%
%
 For example, with $M=2^{7}$ we have $\max_{\l \ne m} \left|\Re \left\{C_{\l,m}\right\}\right|=0.045$ and the maximum penalty is  $\Delta_{\max}=
 0.2$ dB.}
%
%
%
%
%
%
\subsection{Discrete-time description}
For a simple receiver implementation it has been proposed to sample the received signal at chip rate, i.e., every $T_c= \T/M=1/B$ seconds \cite{Sfo:13}. In this case  we have in the interval $[0,\T[$ the samples 
\begin{align}
x(k T_c;\a)&= \exp\left\{j 2 \pi B \frac{k \T}{M} \left[\frac{\a}{M} -\frac{1}{2}+\frac{B k \T}{2 M^2} \right. \right. \nonumber\\
&\left.\left. \hspace{4cm} - u\left(k \frac{\T}{M}-\frac{M-a}{B}\right) \right]\right\} \nonumber\\
&=\exp\left\{j 2 \pi k \left[\frac{a}{M} -\frac{1}{2} +\frac{k}{2\,M} - u\left(k \frac{\T}{M}-\frac{M-a}{B}\right) \right]\right\}  \nonumber\\
&=\exp\left\{j 2 \pi k \left(\frac{a}{M} -\frac{1}{2} +\frac{k}{2\,M} \right)\right\},\, k = 0, 1, \ldots , M-1
\label{eq:x2sampled}
\end{align}
where the last equality is due to the fact that $2 \pi k \,u(\cdot)$ is always an integer multiple of $2 \pi$. This observation allows to avoid the modulus operation in the discrete-time description. 
Then, from \eqref{eq:x2sampled} we have immediately the following property about the orthogonality of the discrete-time waveforms. 
\begin{property}
The discrete-time signals $x(k T_c;\a)$ are orthogonal in the sense that
\begin{align}
\frac{1}{M} \sum_{k=0}^{M-1} x(k T_c; \l) x^*(k T_c; m) = \delta_{\l-m}
\label{eq:xortho}
\end{align}
\end{property}
\begin{proof}
From \eqref{eq:x2sampled} we have 
\begin{align*}
\frac{1}{M} \sum_{k=0}^{M-1} x(k T_c; \l) x^*(k T_c; m) &= \frac{1}{M} \sum_{k=0}^{M-1} e^{j 2 \pi k \left(\frac{\l-m}{M} \right)}\\  &=\delta_{\l-m}
\end{align*}
\end{proof}
As observed in \cite{Sfo:13,Van:17}, once we have $x(k T_c;\a)$ we can compute the twisted (dechirped) vector $\tilde{\bf x}$ with elements
\begin{align}
\tilde{x}_k=\tilde{x}(k T_c;\a)&= x(k T_c;\a) e^{-j 2 \pi \frac{k^2}{2\,M}+j \pi k} \,.
\label{eq:x2st}
\end{align}
Now, substituting \eqref{eq:x2sampled} in \eqref{eq:x2st}, we see that 
\begin{align}
\tilde{x}_k=e^{j 2 \pi k \frac{a}{M}}, \qquad k \in \{0, 1, \ldots , M-1\} 
\label{eq:x2st2}
\end{align}
which can be interpreted as a discrete-time complex sinusoid at frequency $a$. It follows that its Discrete Fourier Transform gives the vector ${\bf X}=\text{DFT}(\tilde{\bf x})$ with elements 
\begin{align}
X_q&=\sum_{k=0}^{M-1} \tilde{x}(k T_c;\a) e^{-j 2 \pi k q/M} = \sum_{k=0}^{M-1} e^{-j 2 \pi k (q-a)/M} \nonumber\\
&= M \delta_{q-a}, \qquad q \in \{0, 1, \ldots , M-1\} \,.
\label{eq:x2sampledtwistedDFT}
\end{align}
Therefore, the DFT of the twisted signal \eqref{eq:x2st} has only one non-zero element in the position of the modulating symbol $a$. 
This means that a possible way to implement a demodulator is to compute the dechirped vector \eqref{eq:x2st}, and decide based on its DFT. 
\begin{remark}
One could think now that working in the discrete-time domain we can achieve the performance of orthogonal modulations. However, this is not exactly the case, since, as will be shown in the next section, the bandwidth of the signal in \eqref{eq:itLORA} is larger than $B$. Therefore, filtering  over a bandwidth $B$ will distort the signal, and the resulting samples will not be like in \eqref{eq:x2sampled}. As a consequence, they will not obey the orthogonality condition in \eqref{eq:xortho}. To avoid distortion, in general a bandwidth larger than $B$  should be kept before sampling. In the presence of AWGN, this will produce an increase in the noise power and correlation between noise samples with respect to an orthogonal modulation. However, for large $M$ the bandwidth of the signal stays approximately into a bandwidth $B$ (see next section and Table~\ref{table:bw}), and therefore it is possible to implement a receiver based on sampling at rate $B$, dechirping, and looking for the maximum of the DFT. This is consistent with the observation that for large $M$ the modulation is approximately orthogonal (see Property~\ref{pro:xcorr}). 
\end{remark}
%
\section{ Spectral Analysis of the LoRa modulation}\label{sec:psd}
{ In this section, the power spectrum of the LoRa modulation is analytically derived in closed form in terms of Fresnel functions, or   through the discrete Fourier transform. Then, it is shown that the modulated signal has a discrete spectrum containing a fraction $1/M$ of the overall signal power.}
\subsection{ Power Spectrum of LoRa Modulated Signals}
Let us consider a source that emits a sequence of \acl{i.i.d.} discrete \aclp{r.v.} $\an$ with probability
\begin{align*}
&\mathbb{P}\{ \an=\ell \}=\frac{1}{M}, &\forall \ell\in \{0,1,\cdots,M-1\}.
\end{align*}
%
From \eqref{eq:itLORA} the modulator output can be represented by the stochastic process
\begin{equation}\label{eq.modulatedsequence}
I(t)=\sum_n x(t- n\T;\an) g_{\T}(t-n\T) 
\end{equation}
where the random signal  $x(t;\cdot)$ can take values in the set $\{x(t;\ell)\}_{\ell=0}^{M-1}$ of finite energy deterministic waveforms. 
The \acl{PSD} of the random process $I(t)$ can be written as the sum of a continuous and a discrete parts
\begin{align}\label{eq.psdgeneral}
G_{I}(f)=G_{I}^{\textrm{c}}(f)+G_{I}^{\textrm{d}}(f) \,.
\end{align}
The expressions of the continuous and discrete spectra in \eqref{eq.psdgeneral} can be found by using for the random process \eqref{eq.modulatedsequence} the frequency domain analysis of randomly modulated signals (see e.g. \cite{Ben:99,Proakis:95}), obtaining 
\begin{align}
G_{I}^{\textrm{c}}(f)&=\frac{1}{\T\,M}\left[\sum_{\ell=0}^{M-1} \left|X(f;\ell)  \right|^{2} - \frac{1}{M} \left|\sum_{\ell=0}^{M-1} X(f;\ell)      \right|^{2}                                \right] \label{eq:gc} \\
G_{I}^{\textrm{d}}(f)&=\frac{1}{\T^{2}\,M^{2}} \sum_{n=-\infty}^{\infty} \left|\sum_{\ell=0}^{M-1} X\left(n\frac{B}{M};\ell\right)\right|^{2} \, \delta\left(f-n\frac{B}{M}\right)                             \label{eq:gd}
\end{align}
where $\{X(f;\ell)\}_{\ell=0}^{M-1}$ are the Fourier transforms of the waveforms $\{x(t;\ell)\}_{\ell=0}^{M-1}$ given in \eqref{eq:x2}. 
%
%
%
%

The spectrum can be derived analytically by expressing the Fourier transforms $X(f;\ell)$ in terms of Fresnel functions. More precisely, we have
\begin{align}
\!\!\!X(f;\ell)&\!\!=\!\! \int_{0}^{\T}\!\!\! x(t;\ell) e^{-j 2\pi f t}     \, dt 
\!=\!\!\!\int_{0}^{\tau_{\ell}}\!\!\!   e^{j 2 \pi \left[B t (\frac{\ell}{M} \!-\!\frac{1}{2})+\! \frac{B^2}{2M} t^2 \right]} e^{-j 2\pi f t}  dt \nonumber \\ 
&+\int_{\tau_{\ell}}^{\T}   e^{j 2 \pi \left[B t (\frac{\ell}{M} -\frac{3}{2})+\frac{B^2}{2\,M}\,t^2 \right]}\, e^{-j 2\pi f t} \, dt. \label{eq.sinfint}
\end{align}
Let us define the function 
\begin{align}
W(a; b; t_1; t_2)&= \int_{t_1}^{t_2}  \exp \left({j 2 \pi \left[a\,t+b\, t^{2}\right]}\right)\, dt \label{eq:Wdef}
\end{align}
that can be expressed in terms of the Fresnel functions as 
\begin{align}
W(a; b; t_1; t_2)&=  \frac{1}{2\sqrt{b}} 
\,e^{-j 2 \pi \frac{a^2}{4\,b}} \left[K\left(2\sqrt{b}\,\left(t_{2}+\frac{a}{2\,b}\right) \right) - \right. \nonumber \\
& \left. K\left(2\sqrt{b}\,\,\left(t_{1}+\frac{a}{2\,b}\right) \right)  \right] \label{eq.generalint}
\end{align}
where $K(x)\triangleq C(x)+j\,S(x)$. 
Then, the Fourier transform of the waveforms can be written analytically as 
\begin{align}\label{eq:Xf}
X(f;\ell)&= W\left(B \left(\frac{\ell}{M} -\frac{1}{2}\right)-f; \frac{B^2}{2\,M}; 0; \frac{M-\ell}{B}\right)+ \nonumber \\
&W\left(B \left(\frac{\ell}{M} -\frac{3}{2}\right)-f; \frac{B^2}{2\,M}; \frac{M-\ell}{B};\frac{M}{B}\right) \,
\end{align}
that used in \eqref{eq:gc} and \eqref{eq:gd} gives the signal spectrum. 
%
%

An alternative to the use of the Fresnel functions consists in the standard Discrete Fourier Transform approach,  
where we take $N$ samples of $x(t;\ell)$ over the time interval $[0,\T[$ in a vector ${\bf{x}}(\ell)=\{x(0;\ell), x(\Delta_t;\ell), \cdots, x((N-1)\Delta_t;\ell)\}$, with step $\Delta_t=\T/N$.   
Then, the vector ${\bf{X}}(\ell)=\Delta_t \, \text{DFT}({\bf{x}}(\ell))$ gives 
 the samples with frequency step $\Delta_f=1/\T= B/M$ of the periodic repetition $\sum_k X(f-k F;\ell)$, where $F=N/\T= N B / M$.  For sufficiently large $N$  the effect of aliasing is negligible
 , so that the elements of ${\bf{X}}(\ell)$ are essentially the samples of $X(f;\ell)$ with step $\Delta_f$. 
 For the discrete spectrum this frequency step is exactly what is needed in \eqref{eq:gd}.  
If a finer resolution in frequency is needed (for the continuous spectrum in \eqref{eq:gc}) we have to zero-pad the vector ${\bf{x}}(\ell)$ before taking the DFT. For example, if we add $(k-1)N$ zeros to ${\bf{x}}(\ell)$ the frequency step is  $\Delta_f=1/k\T= B/kM$. 
\subsection{ Total Power of the Discrete spectrum}
Lines in the spectrum indicates the presence of a non-zero mean value of the signal, which does not carry information. 
The following property quantifies the power of this mean value with respect to the overall signal power. 
\begin{property}\label{property:discretepowerratio}
	The  total power of the discrete spectrum for the LoRa modulation 
\begin{align*}
P_\textrm{d} &= \int_{-\infty}^{\infty} G_{I}^{\textrm{d}}(f)\, df = \frac{1}{\T^{2}\,M^{2}} \sum_{n=-\infty}^{\infty} \left|\sum_{\ell=0}^{M-1} X\left(n\frac{B}{M};\ell\right)\right|^{2} \,
\end{align*}
	is exactly a fraction $1/M$ of the overall signal power.
\end{property}
\begin{proof}
The discrete spectrum in \eqref{eq:gd} is due to the mean value of the signal
\begin{equation*}
\EX{I(t)}=\sum_n \EX{x(t- n\T;\an)} g_{\T}(t-n\T) \,.
\end{equation*}
This mean value is not zero, implying that there are lines in the spectrum\cite{Proakis:95,Ben:99}. More precisely, since the modulation is memoryless, we have for $0 \leq t < \T$
\begin{align*}
&\EX {x(t;A_0)} = \frac{1}{M} \sum_{\ell=0}^{M-1} x(t;\ell) 
=\frac{1}{M} \sum_{\ell=0}^{M-1} \sum_{k=0}^{M-1} x(t;\ell)      \\
&\phantom{=}\times g_{T_{c}}\left(t-k\,T_{c}\right)=\frac{1}{M} \sum_{k=0}^{M-1}  g_{T_{c}}\left(t-k\,T_{c}\right) \sum_{\ell=0}^{M-1} x(t;\ell)\\
&=\frac{1}{M}\left\{g_{T_{c}}(t)\sum_{\ell=0}^{M-1} x(t;\ell)+  \sum_{k=1}^{M-1}  g_{T_{c}}\left(t-k\,T_{c}\right) \sum_{\ell=0}^{M-1} x(t;\ell)                      \right\}
\end{align*}
where $T_{c}=1/B$ is the chip rate. From \eqref{eq:x2} we have
\begin{align*}
&\EX {x(t;A_0)}  =\frac{1}{M} e^{j 2 \pi \frac{B}{2 \T}\, t^2}\left\{ g_{T_{c}}(t)\sum_{\ell=0}^{M-1} e^{j 2 \pi \frac{B}{M}\ell\, t}\right.+ \\
&\left. \sum_{k=1}^{M-1} g_{T_{c}}\left(t-k\,T_{c}\right) \left[\sum_{\ell=0}^{M-k-1}e^{j 2 \pi \frac{B}{M}\ell\, t}+ \sum_{\ell=M-k}^{M-1}e^{j 2 \pi \frac{B}{M}\ell\, t} e^{-j 2 \pi B t}        \right]                        \right\}\\
&=\frac{1}{M} e^{j 2 \pi \frac{B}{2 \T}\, t^2}\left\{g_{T_{c}}(t) \frac{1-e^{j  2 \pi B  t}}{1-e^{j  2 \pi B t/M}}+\sum_{k=1}^{M-1} g_{T_{c}}\left(t-k\,T_{c}\right)\right.\\
&\phantom{=}\left.\times e^{j 2 \pi B (M-k)t/M}\frac{e^{-j  2 \pi B t}-1}{1-e^{j  2 \pi B t/M}}\right\}.
\end{align*}
After some manipulation we get
\begin{align*}
\EX {x(t;A_0)}&=\frac{1}{M}e^{j \frac{\pi B t}{M} (B t - 1)} \frac{\sin \left(\pi B t\right)}{\sin \left(\pi B t /M\right)} \\ 
&\phantom{=}\times  \sum_{k=0}^{M-1}g_{T_{c}}\left(t-k\,T_{c}\right)\, e^{-j 2 \pi B k t /M}.
\end{align*}
%
%
The absolute value of the mean is therefore
\begin{align*}
|\EX{x(t;A_0)}| &= \frac{1}{M} \left|\frac{\sin \left(\pi B t\right)}{\sin \left(\pi B t /M\right)} \right|, && 0 \leq t < \T \,.
\end{align*}
Now, recalling the following integral for $m$ integer \cite[p. 396]{GraRyz:B07}
\begin{align*}
\int_{0}^{\pi/2} \left(\frac{\sin m x}{\sin x} \right)^{2}dx=\frac{\pi}{2}
\end{align*}
we get the power of the discrete spectrum as 
\begin{align}\label{eq.intabs2Eit}
P_\textrm{d}=\frac{1}{\T} \int_0^{\T} |\EX{x(t;A_0)}|^2 \, dt = \frac{1}{M}\,.
\end{align}
Therefore, there are lines in the spectrum of the LoRa modulation, and  the power of this discrete spectrum is a fraction $1/M$ of the overall power. 
\end{proof}
\section{Numerical Results}
%
%
We first show in Fig.~\ref{fig:psd} the two-sided power spectrum of the complex envelope  for LoRa modulated signals as a function on the normalized frequency $f/B$, with various spreading factors, i.e., $\SF \in \{3,7,10,12\}$. 
Since $G_I(-f)=G_I(f)$ 
 we just show $G_I(f)$ for $f\geq 0$.  
\begin{figure}
	\centering
	\begin{subfigure}[b]{0.99 \linewidth}
		\includegraphics[width=0.99\linewidth]{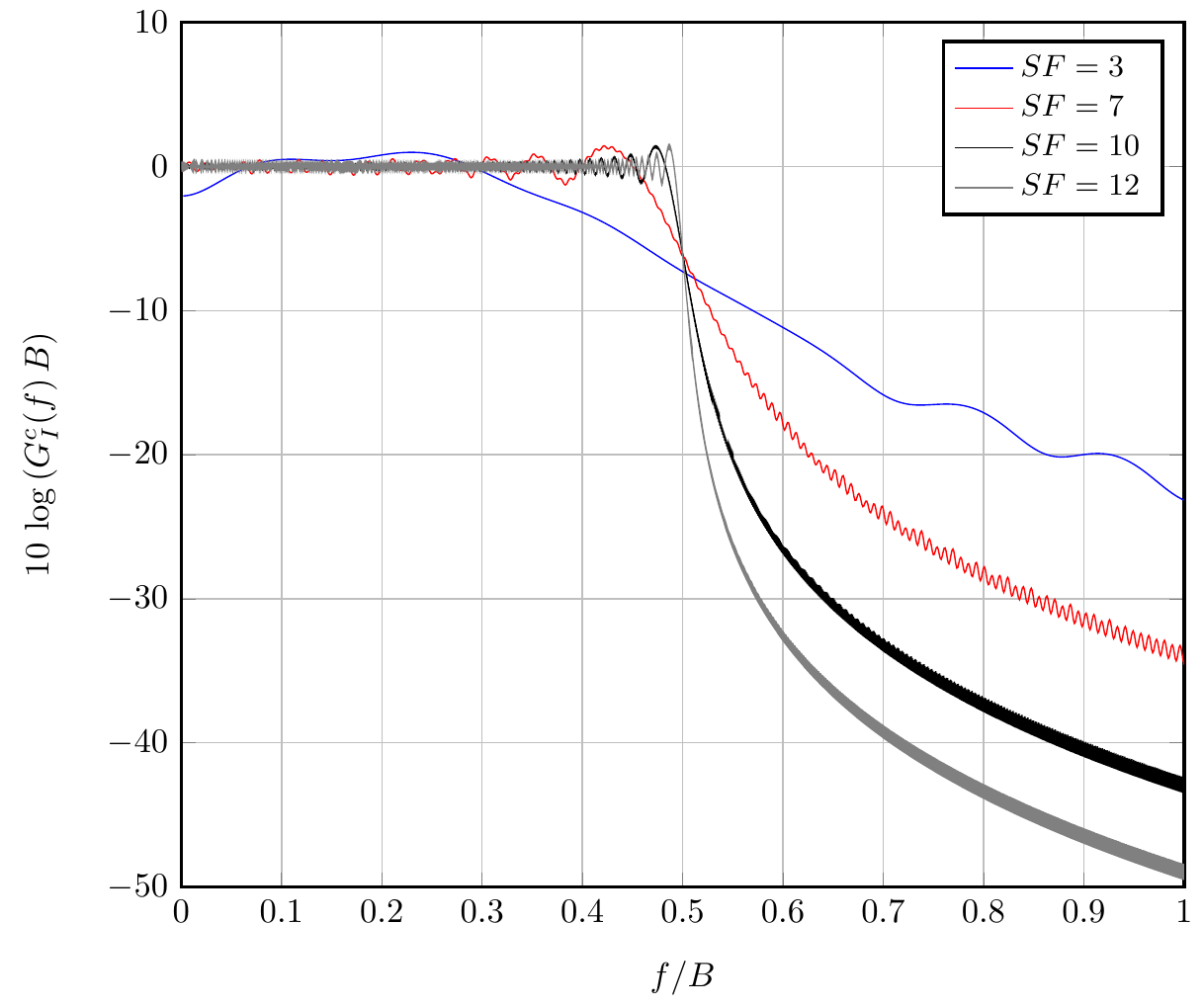}
		\caption{The continuous part of the spectrum}\label{fig:PSdCont}
	\end{subfigure}

	\medskip
		
	\begin{subfigure}[b]{0.99 \linewidth}
		\includegraphics[width=0.99\linewidth]{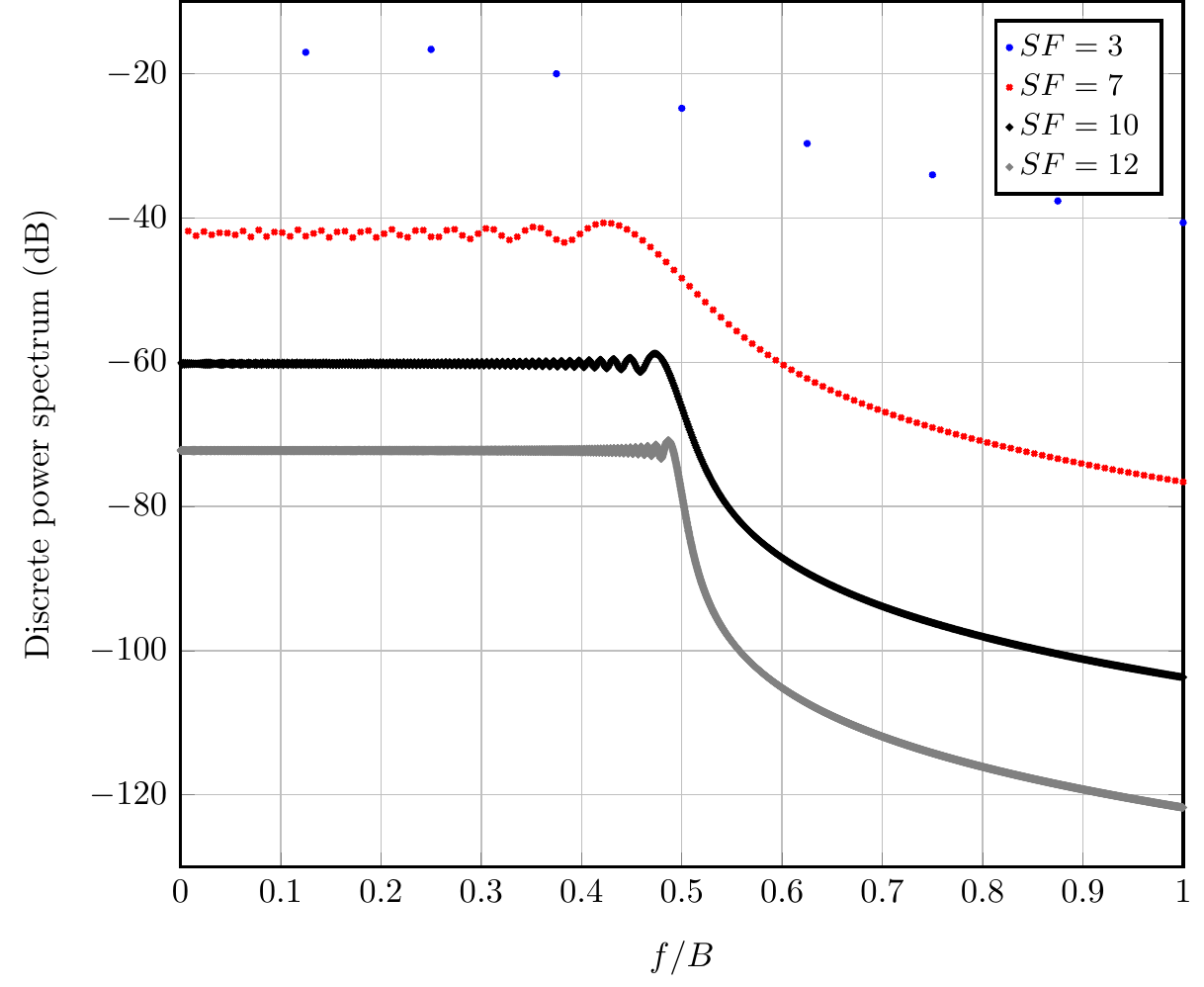}
		\caption{The discrete spectrum}\label{fig:PSDDisc}
	\end{subfigure}
	\caption{The continuous and discrete spectrum of the complex envelope for LoRa modulation, $M=2^\SF$, with $\SF \in \{3,7,10,12\}$.}
	\label{fig:psd}
\end{figure}
In the figure we report both the normalized power spectral density, $10\,\log_{10} G^{c}_{I}(f)\, B$, and the discrete part of the spectrum. For the latter we report the power $\left|\sum_{\ell=0}^{M-1} X\left(n {B}/{M};\ell\right)\right|^{2} / {\T^{2}\,M^{2}}$ at frequency $n B/M$, as given in \eqref{eq:gd}. The sum of the power of all lines in the discrete spectrum is equal to $1/M$, as proved in Property~\ref{property:discretepowerratio}. 
For example, with $\SF=3$ we have $M=8$ and thus $1/M=12.5\%$ of the signal power is contained in the discrete spectrum.  
We can see that the power spectrum becomes more compact for increasing $M$, so that most of the power for the complex envelope is contained between $-B/2$ and $B/2$, or, in other words, that the modulated signal bandwidth is close to $B$ for large $M$. 

To better quantify this effect, we report in Table~\ref{table:bw} the bandwidth $B_{99}$ centered on $f_0$ containing $99\%$ of the power for different spreading factors. It can be seen that, while for $M \ge 2^7$ almost all of the signal is contained in a bandwidth $B$, considering just a bandwidth $B$ for smaller spreading factors will leave out a part of the signal, therefore distorting the signal. Moreover, as noted in Section~\ref{sec:signalmodel} and Section~\ref{sec:psd}, the spectral efficiency, the maximum real cross-correlation, and the power of the discrete spectrum decrease for increasing $M$.  

\begin{figure}
	\centering
	\begin{subfigure}[b]{0.99 \linewidth}
		\includegraphics[width=0.99\linewidth]{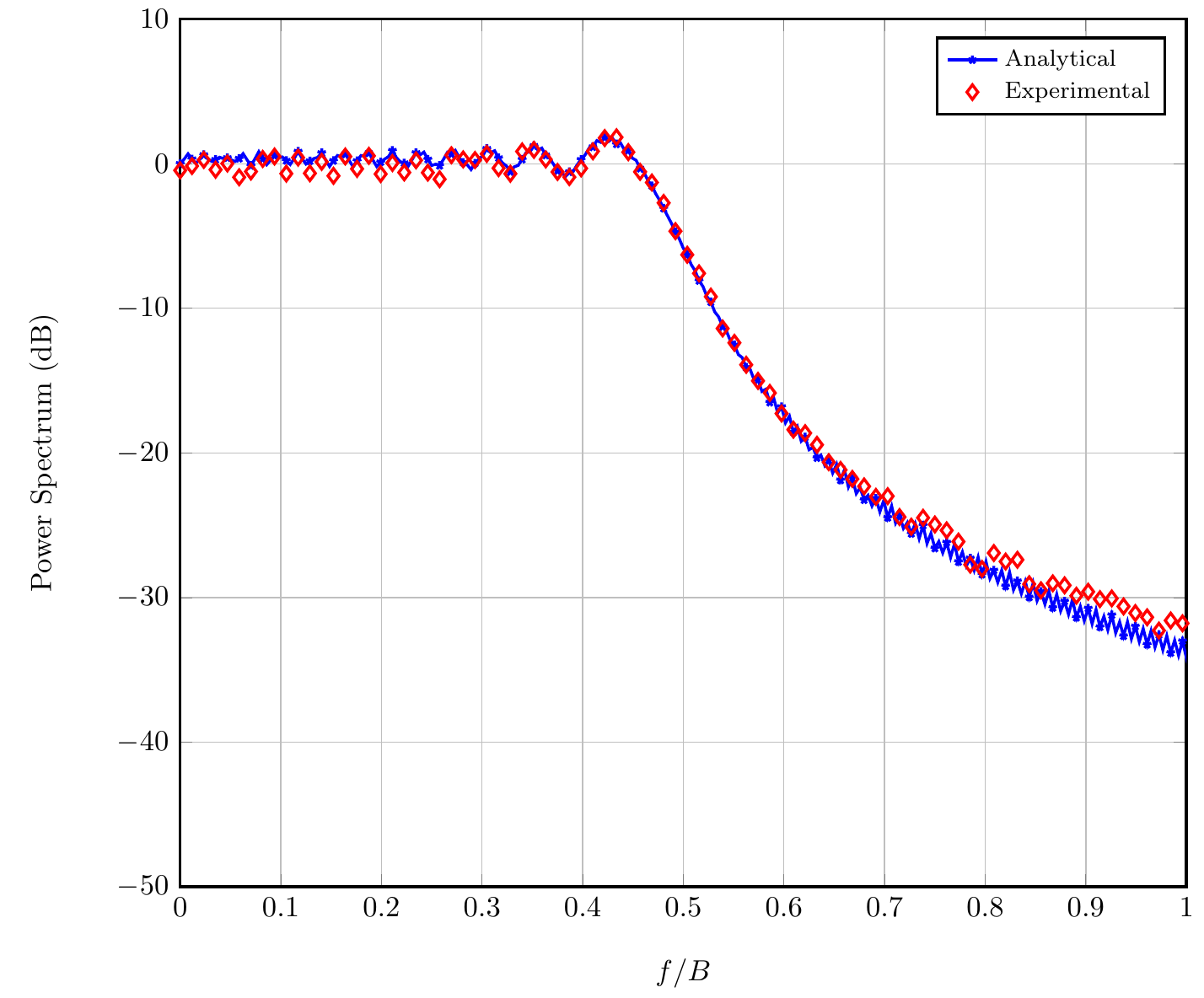}
		\caption{$\SF=7$}
	\end{subfigure}

	\medskip
	
	\centering	
	\begin{subfigure}[b]{0.99 \linewidth}
		\includegraphics[width=0.99\linewidth]{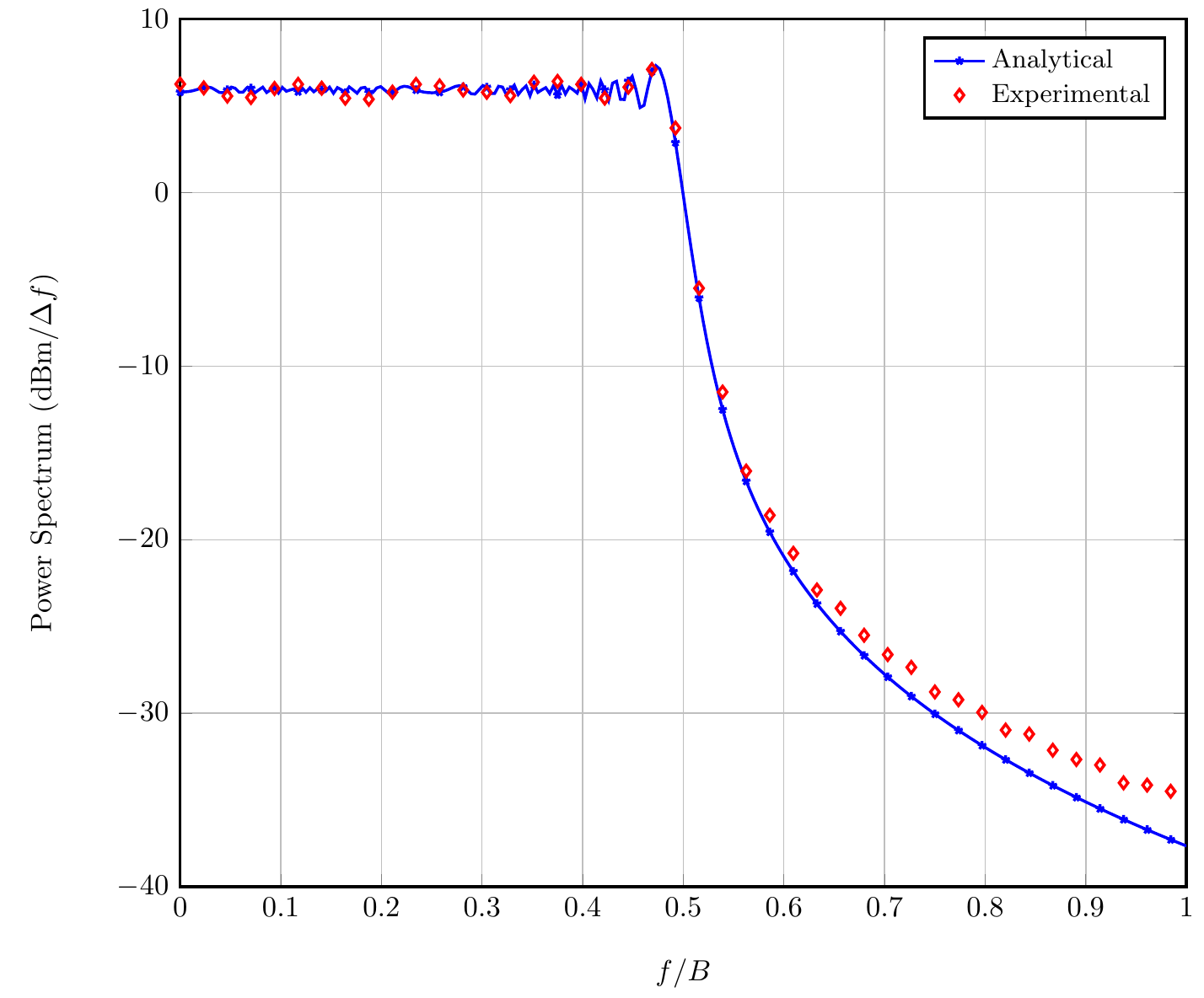}
		\caption{$\SF=10$}
	\end{subfigure}
	\caption{The power spectrum  of the complex envelope for LoRa modulation using the analytical expressions and the experimental data, for $M=2^\SF$, $\SF \in \{7,10\}$, $B=125$~kHz, $\Delta f=B/256$, and $P_{\text s}=27$~dBm.}
	\label{fig:figpowerspectrumexactiq}
\end{figure}
In Fig.~\ref{fig:figpowerspectrumexactiq}, we compare the derived analytical power spectrum with that obtained from the IQ samples of a commercially available LoRa transceiver \cite{IQsemtech}. 
{
More precisely, IQ samples are provided for LoRa modulated waveforms, which have been created with a randomly generated payload of $16$ bytes.
The waveforms are obtained for $B=125~$kHz with sample rate $f_{s}=4\,B$ \cite{IQsemtech}.} The frequency range of interest is divided into several bins with width $\Delta f=B/256$, and the power within each bin is computed either analytically via \eqref{eq:gc} and \eqref{eq:gd}, or through spectral estimation by implementing the Welch's method on the experimental data\cite{Welch:67}. It is noticed that the estimated spectrum agrees well with the analytical expression. We can also observe that the tail of the estimated spectrum is slightly higher than the analytical; this is because the experimental samples have been taken at $f_{s}=4\,B$, not large enough to completely eliminate frequency aliasing.

\begin{figure}
	\centering
	\begin{subfigure}[b]{0.99 \linewidth}
		\includegraphics[width=0.99\linewidth]{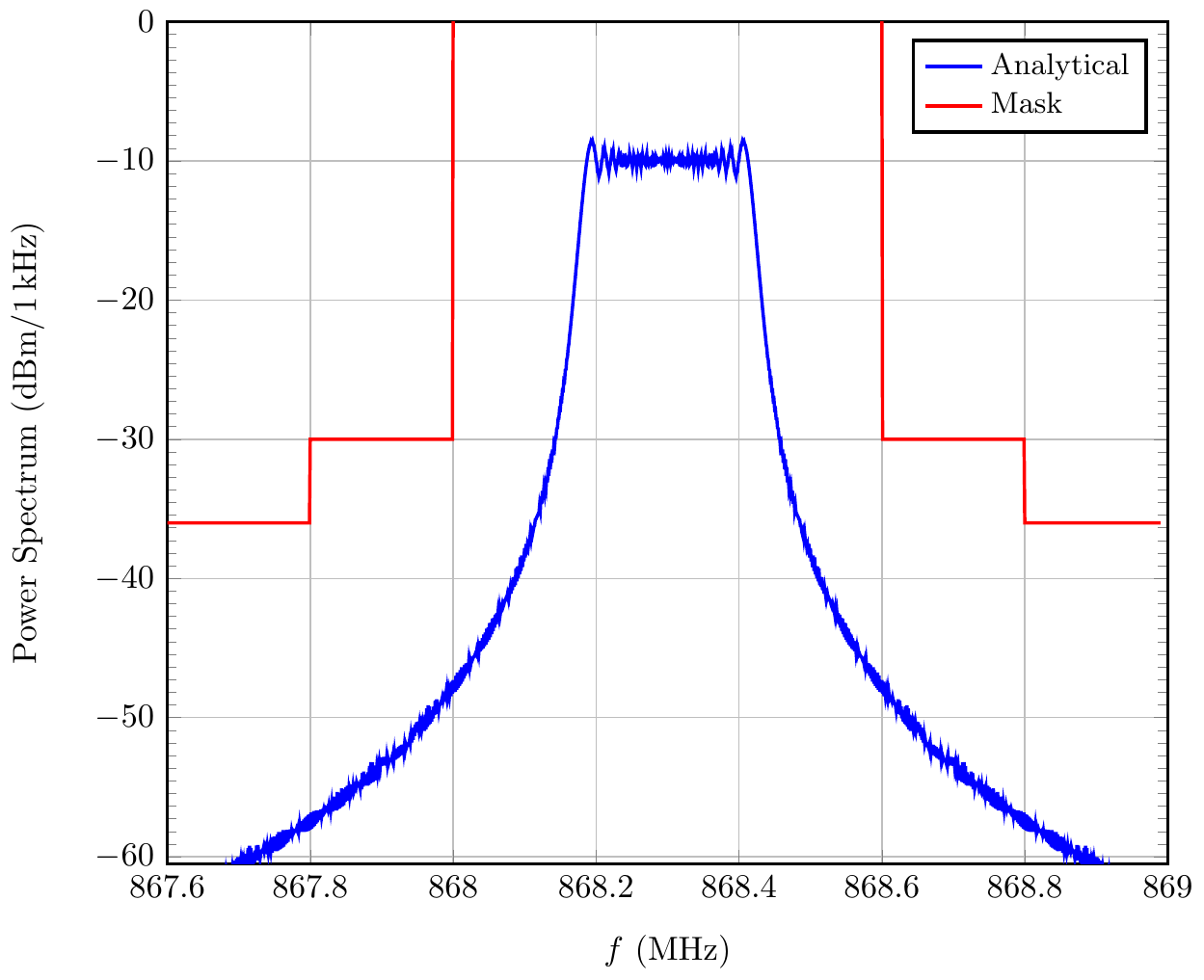}
		\caption{One channel, $B=250$~kHz.}
		\label{fig:spectrumwithmaskone}
	\end{subfigure}

	\medskip
		
	\begin{subfigure}[b]{0.99 \linewidth}
		\includegraphics[width=0.99\linewidth]{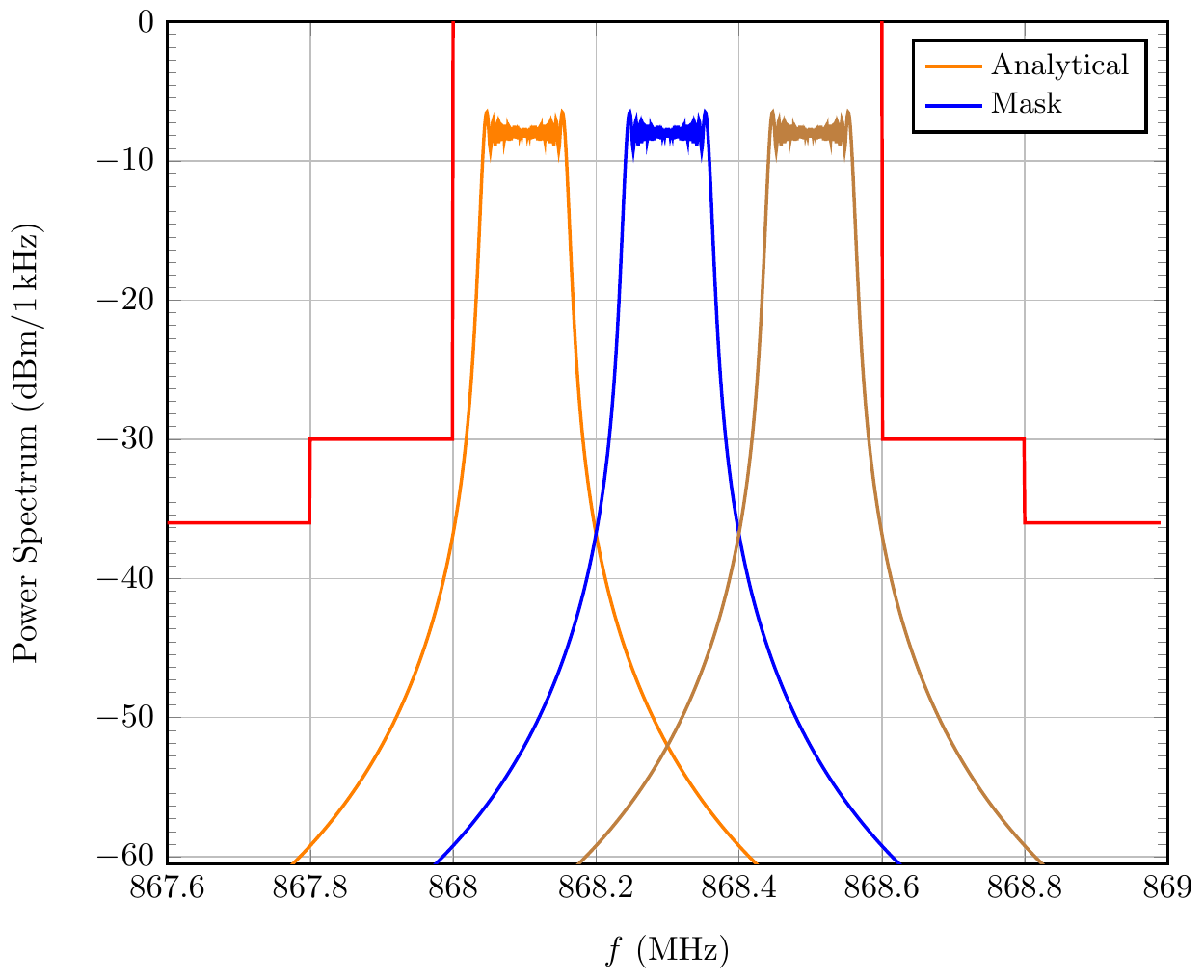}
		\caption{Three channels, $B=125$~kHz.}
		\label{fig:spectrumwithmasthreek}
	\end{subfigure}
	\caption{The one-sided power spectrum for LoRa modulated passband signals using the analytical expressions, compared with the mask from the ETSI regulation in the G1 sub-band, for $M=2^7$, $\Delta f=1$~kHz, and $P_{\text s}=14$~dBm. \\
	a) One channel with center frequency $868.3$~MHz for $B=250$~kHz.
	b) Three channels with center frequencies $868.1$~MHz, $868.3$~MHz, and $868.5$~MHz for $B=125$~kHz.
	}
	\label{fig:spectrumwithmask}
\end{figure}
{ Finally, we investigate the LoRa spectrum along with the ETSI regulations for out-of-band emissions \cite[7.7.1]{ETSI:12}. Since LoRa is a chirp spread spectrum technique, it is governed by the regulations for ISM bands that support wideband modulation \cite[Table~5]{ETSI:12}. For example, we consider the G1 sub-band spanning from $868$\,MHz  to $868.6$\,MHz\cite[Fig.~7]{ETSI:12}. There are two possibilities for using LoRa in this sub-band:
	\begin{itemize}
		\item using a single channel  with center frequency $868.3$\,MHz for  $B=250$\,kHz;
		\item using three channels with center frequencies $868.1$, $868.3$, and $868.5$\,MHz for $B=125$\,kHz.
	\end{itemize}  
	In Fig.~\ref{fig:spectrumwithmask} we report, for the two cases above, the one-sided power spectrum  calculated analytically with bin width (i.e., resolution bandwidth) $\Delta f=1$\,kHz, and $P_{\text s}=14$\,dBm, i.e., the maximum allowed transmission power. The spectrum is compared with the spectral mask for the G1 sub-band.	It can be noticed that the spectrum meets the regulations of the maximum power limits for adjacent band emissions at the G1 sub-band. The same method can be used to examine the LoRa compliance for various ISM bands, spreading factors, and bandwidths, according to other regional regulations.
   } 
\section{Conclusions}
In this paper we investigated the spectral characteristics of the LoRa $M$-ary modulation, deriving the analytical expression of the spectrum, and comparing it  with experimental data { and with the spectral limit masks for the ISM bands}.  We found that there are lines in the spectrum, containing a fraction $1/M$ of the overall power, and that the occupied bandwidth is in general larger than the deviation $B$. We also derived the waveform cross-correlation function, proving that the LoRa waveforms can be considered  orthogonal only for asymptotically large $M$.    
\bibliographystyle{IEEEtran}
\enlargethispage{-6.8cm}
\bibliography{IEEEabrv,LoRa_biblio}
\end{document}